# Constructing quasicrystalline lattices

O. V. Konevtsova and S. B. Rochal

*Physical Faculty, Southern Federal University, 5 Zorge Street, 344090 Rostov-on-Don, Russia*

A new method to construct quasicrystalline lattices is proposed. It is based on Landau crystallization theory. Like well-known cut and projection methods our approach deals with N dimensional crystallography, but we don't need any conception similar to an acceptance domain or atomic surfaces. The selection of nodes, included into the lattice, is based on simple rules provided their local order in a real space.

## 1. Introduction

Quasicrystals (QC's), metallic alloys with surprising properties were found in 1984 [1]. Their diffraction patterns were similar to those of ordinary crystals: there were bright and regularly spaced reflections. Nevertheless, their patterns exhibited an icosahedral symmetry prohibited in crystals due to purely geometrical reasons. This strange discovery induced a great interest in the fields of condensed matter physics and crystallography. First theoretical description of QC's was published in several months [2]. It was based on a space covering by two tiles, proposed earlier by Penrose [3]. The tiles were rhombuses with equal sides but different angles $2\pi/5$ and $\pi/5$, respectively. Penrose tiling has combined high rotational symmetry (forbidden in crystals) with long-range order of non-translation type. Two - tiles model explained qualitatively the symmetry of diffraction patterns and became dominating for several next years. For construction of different tilings several approaches were used: generalized dual method [4-5], method, based on self-similarity [6], cut and projection methods [7-12]. To relate the tilings obtained with real structures the tiles were considered as different unit cells decorated by appropriate atoms [9-10,13]. Later, the concept of two different unit cells was practically rejected.

QC's are defined now as aperiodic structures with a long-range order and noncrystallographic rotational symmetry. The conception of N-dimensional crystallography proposed by Wolf for modulated phase in 1974 [14] appeared to be suitable for QC's too. Within such approach their structures are described by N-dimensional space groups spanning corresponding N-dimensional spaces periodically decorated by atomic surfaces (AS's) [15]. Intersection of these surfaces with a cut plane generates atomic positions in the real structure.

Less rigorous from a crystallographic point of view, but equivalent and simpler description of real QC structures was proposed by Burkov [16]. As an example, decagonal structure $Al_{65}Cu_{20}Co_{15}$ was considered [16]. According to Burkov, the QC structure is formed from equivalent building blocks (clusters), situated in the nodes of a quasilattice (QL). In his original work, Burkov used Penrose tiling [16] as the QL. Unlike unit cells in crystals, the clusters in Burkov's model are overlapped, but the number of different overlapping is limited. For example, in the $Al_{65}Cu_{20}Co_{15}$ structure only two types of overlapping exist.

Later, the cluster approach became a basis of many structural models of QC's [17-21]. Burkov's model for $Al_{65}Cu_{20}Co_{15}$ structure was improved by Yamamato [19], who proposed to decorate another QL (see its description in the next section) by the same clusters. On the basis of Yamamato`s model many decagonal structures were interpreted, for example, that of $Al_{65}Cu_{15}Co_{20}$ alloy [21].

Thus, a structural organization of QS's is similar to that of ordinary crystals. Burkov's cluster is analogous to a crystalline unit cell, and the QL, which plays an important role in the organization of quasicrystal order, is similar to a crystalline Bravais lattice. The main purpose of the present work is the further development of the QL theory. We propose a new method to construct QL's. Our approach uses N-dimensional crystallography, but, unlike cut and projection methods, it is based on Landau theory of crystallization. Crystallographic fundamentals of the method proposed are considered in the next section. Section 3 explains how our method is based on Landau theory of crystallization.

2. **Pentagonal quasilattices**

Let us start from a crystallography of our approach. Nodes of the QL can be found projecting integer nodes $\{n_i^j\}$, i=0, 1…4 of 5-dimensional phase space $E$. This space is decomposed into three irreducible subspaces. Two of them are 2-dimensional ($E^{\parallel}$ and $E^{\perp}$) and they are spanned by two different vector irreducible representations of $C_{10v}$ symmetry group. Basis functions of the first representation (or projections upon the first subspace) read:

$$\mathbf{r}^j = \sum_{i=0}^{4} n_i^j \mathbf{a}_i \qquad (1)$$

where $\mathbf{a}_i$ are 2D basis vectors, taken in the following form: $\mathbf{a}_i = \left(\cos\left(\frac{i2\pi}{5}\right), \sin\left(\frac{i2\pi}{5}\right)\right)$, i=0,1,2,3,4.

Basis functions of the second representation (or projections upon the second subspace) are:

$$\mathbf{r}_j^\perp = \sum_{i=0}^{4} n_i^j \mathbf{a}_i^\perp \qquad (2)$$

where, $\mathbf{a}_i^\perp = \left(\cos\left(\frac{i6\pi}{5}\right), \sin\left(\frac{i6\pi}{5}\right)\right)$, $i$ =0,1,2,3,4.

And the last 1D subspace corresponds to a totally symmetric representation with basis function:

$$\xi = \sum_{i=0}^{4} n_i^j \qquad (3)$$

Let $n_i^j$ are the integer nodes of space $E$ satisfying condition $\xi$ = *constant*. Then no minimal distance between their projections in space $E^{\|}$ exists. Any vicinity of any point contains an infinite number of positions. Therefore to describe a quasicrystalline order, a selection rule for $n_i^j$ accepted is needed. Selection rules in the cut methods and in the projection one are different. In the projection method $\mathbf{r}_j^\perp + \mathbf{v} \subset \mathbf{S}$, where S is an acceptance domain and $\mathbf{v}$ corresponds to Goldstone degrees of freedom of QC's [10-12]. In the cut method integer nodes of 5D space are decorated by atomic surfaces with the shape S. A node projection is included into the QL provided subspace $E^{\|}$ (1) intersects a corresponding atomic surface. Goldstone degrees of freedom appear in this approach as plane-parallel motions of subspace $E^{\|}$.

We reject any conception similar to the conception of the acceptance domain or of the atomic surfaces. In our approach the node selection is based on the analysis of the local order around this node. Let us characterize this local order by an irreducible translation star Z to the given node from its neighbors. Since all the nodes in the QL have identical $\xi$ (3), the value of $\Delta\xi$ corresponding to translations $\mathbf{Z}_l$ is equal to zero. As the first example, we consider the star of vector $\mathbf{Z}_l$=<1,-1,-1,1,0> composed of 10 vectors $\mathbf{Z}_{i,}$. These vectors (cyclic permutation and inversion of $\mathbf{Z}_l$) are resulted from $\mathbf{Z}_l$ under the action of $C_{10v}$ symmetry group.

In order to formulate our selection rule, we calculate scalar value $l_m$ for a node under consideration and its ten neighbors with respect to the star **Z:**

$$l_m = |\mathbf{r}_m^\perp + \mathbf{v}| \qquad (4)$$

A geometrical sense of Eq. (4) is simple. In the cut method it is the distance between the integer 5d node and plane $E^{\parallel}$. In the projection method $l_m$ is the distance between the node perpendicular projection and the acceptance domain center. Our selection rule is the following. The node is included into the QL, if its neighbors pointed by vectors $\mathbf{Z_i}$ and characterizing by smaller value (4) are absent. Direct application of this algorithm generates well-known pentagonal Penrose QL (Fig. 1).

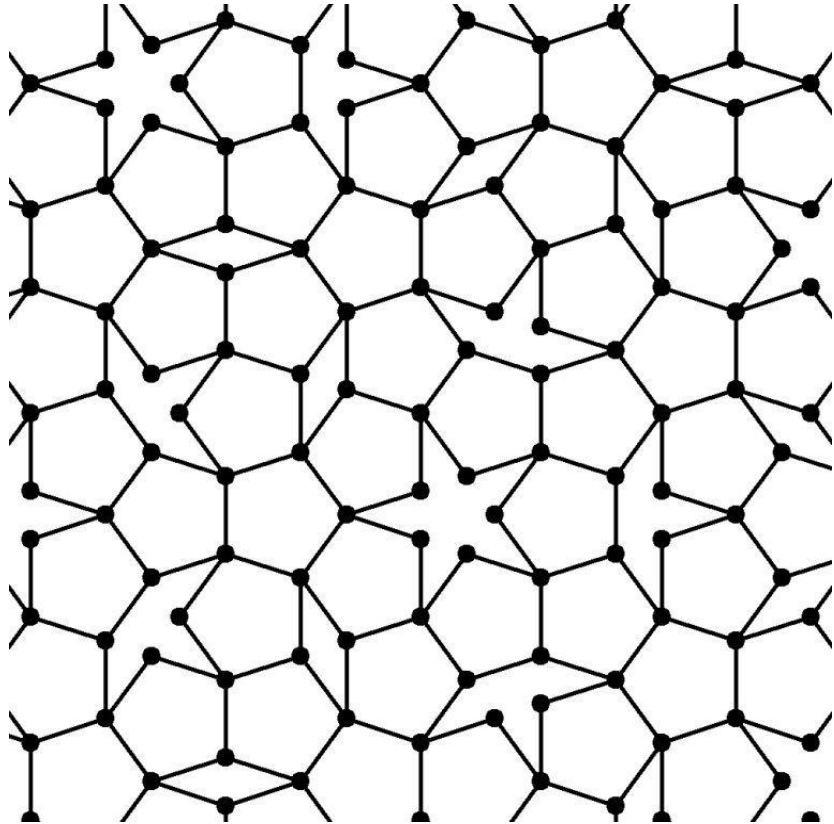

*FIG. 1. The ideal pentagonal tiling formed by regular pentagons, five-armed stars (some of them are truncated), and narrow rhombuses. Circles show nodes of the tiling. Goldstone degree of freedom $\mathbf{v}$ is $<0.2, 0.222>$, $\xi=0$.*

Due to orthogonality of basis functions (1-3) any shift $\{\alpha,\alpha,\alpha,\alpha,\alpha\}$ of the node $\{n_i^{\ j}\}$ does not change its parallel (1) and perpendicular (2) projections. Therefore, two tilings with the difference of $\Delta\xi=5$ are identical in spaces (1) and (2). So, it makes sense to consider five different values of $\xi$ only: $\xi=0,1..4$. However, cases $\xi=2$ and $\xi=3$ differ one from the other by 180° rotation. The same rotations relates $\xi=1$ and $\xi=4$ cases. In the case $\mathbf{v}=0$ and $\xi=1$ or $\xi=4$ the lattice center contains a star. In the case $\xi=0$, $\mathbf{v}=0$ tiling has a global 10-fold symmetry axis. Nevertheless, the same tiles compose all the QL's with different integer $\xi$ values and $\mathbf{v}\neq 0$, since all the structures possess the same free energy. Further consideration of this point will be given in the next section.

To clarify our selection rule let us consider a pair of neighbors translated into the node under selection by two vectors $\mathbf{Z}_k$ and $-\mathbf{Z}_k$. The condition that the accepted node has a smaller value $|\mathbf{r}_m^\perp + \mathbf{v}|$ than these neighbors means that the vector $|\mathbf{r}_m^\perp + \mathbf{v}|$ is projected into a band with a distance between opposite sides equal to the perpendicular projection of $\mathbf{Z}_k$. Intersection of all these bands induced by all $\mathbf{z}_i^\perp$ projections produces a regular decagon, which corresponds exactly to the acceptance domain (see Fig. 2) of pentagonal Penrose tiling. Star Z has been chosen to obtain this well-known structure. To determine $\mathbf{Z}_i$ vectors we used the fact that vectors $\mathbf{z}_i^\parallel$ should be forbidden translations between the QL nodes. As it is known, rhombuses with angle $2\pi/5$ are absent in this tiling. The translations equal to their smaller diagonals are also prohibited. Therefore we simply assumed that these prohibited translations were $\mathbf{z}_i^\parallel$ vectors.

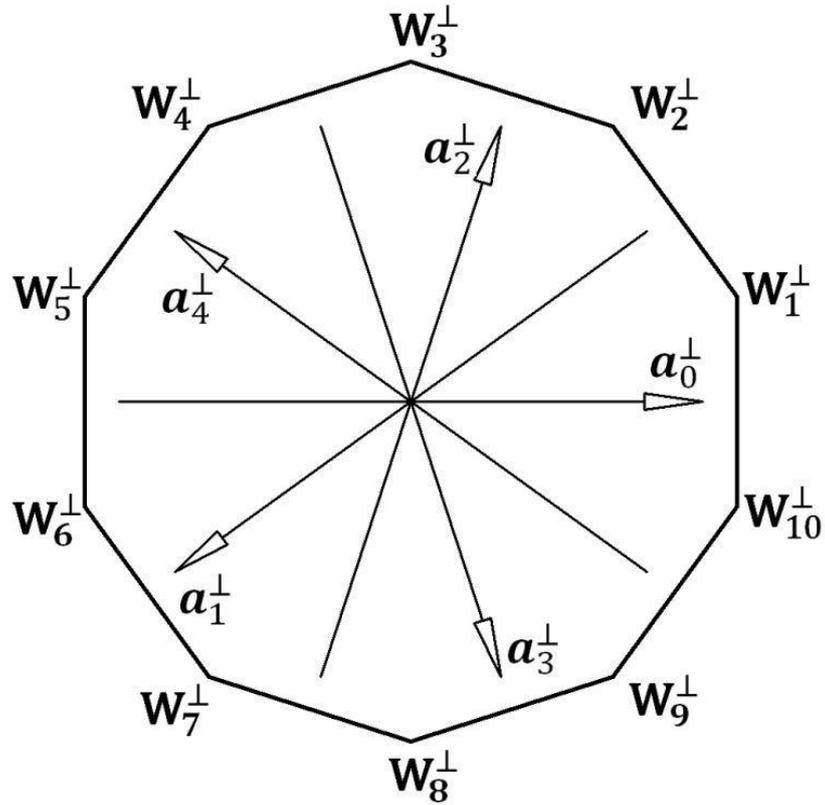

*FIG. 2. Decagonal acceptance domain. Distances between opposite sides are equal to $\mathbf{z}_i^\perp$. Vertexes of the domain coincide with perpendicular projections of the following vectors:* $\mathbf{W}_1^\perp = -\mathbf{W}_6^\perp = \mathbf{a}_2^\perp - \mathbf{a}_4^\perp$, $\mathbf{W}_2^\perp = -\mathbf{W}_7^\perp = \mathbf{a}_0^\perp - \mathbf{a}_3^\perp$, $\mathbf{W}_3^\perp = -\mathbf{W}_8^\perp = \mathbf{a}_4^\perp - \mathbf{a}_1^\perp$, $\mathbf{W}_4^\perp = -\mathbf{W}_9^\perp = \mathbf{a}_2^\perp - \mathbf{a}_0^\perp$, $\mathbf{W}_5^\perp = -\mathbf{W}_{10}^\perp = \mathbf{a}_1^\perp - \mathbf{a}_3^\perp$, *where* $\mathbf{a}_i^\perp = (\cos(6\pi/5), \sin(6\pi/5))$ *basis vectors of* $E^\perp$. $\mathbf{Z}_1 = \mathbf{W}_1 - \mathbf{W}_5 = \mathbf{W}_{10} - \mathbf{W}_6 = \langle 1,-1,-1,1,0\rangle$.

Various QL's can be obtained applying different translation stars and different conditions for modules (4). Further, one more variant of the pentagonal

QL is considered. This case is interesting since a new local selection rule is a condition imposed on the nodes being the nearest neighbors in $E^{//}$ space. The distance between the nearest neighbors is equal to the length of parallel projections of $\mathbf{Z}_i^1$ translations resulted from <1,-1,0,0,0> vector. These projections coincide with smaller diagonals of thin rhombuses (see Fig. 1).

To construct the second QL we apply this nearest neighbor star $\mathbf{Z}^1$ and condition that the node is accepted, if no more than one of its neighbors with respect to star $\mathbf{Z}^1$ has smaller value of module (4). Direct application of this algorithm leads to the QL with another pentagonal local order, represented in Fig.3.

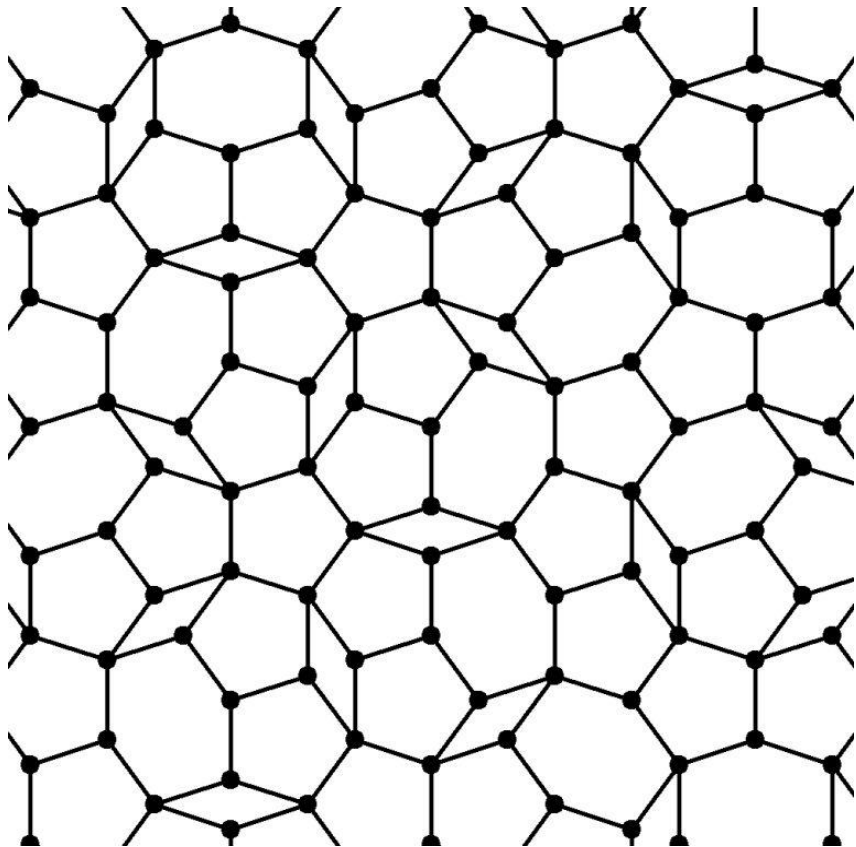

*FIG. 3. The pentagonal tiling consisting of regular pentagons, narrow rhombuses and oblate hexagons. Case of* v=(0.2;0.222), $\xi$=0 *is shown.*

Of course, this QL can be also obtained by the cut and projection methods. But as it is clear from the acceptance domain shape (see Fig. 4) their application is not so convenient.

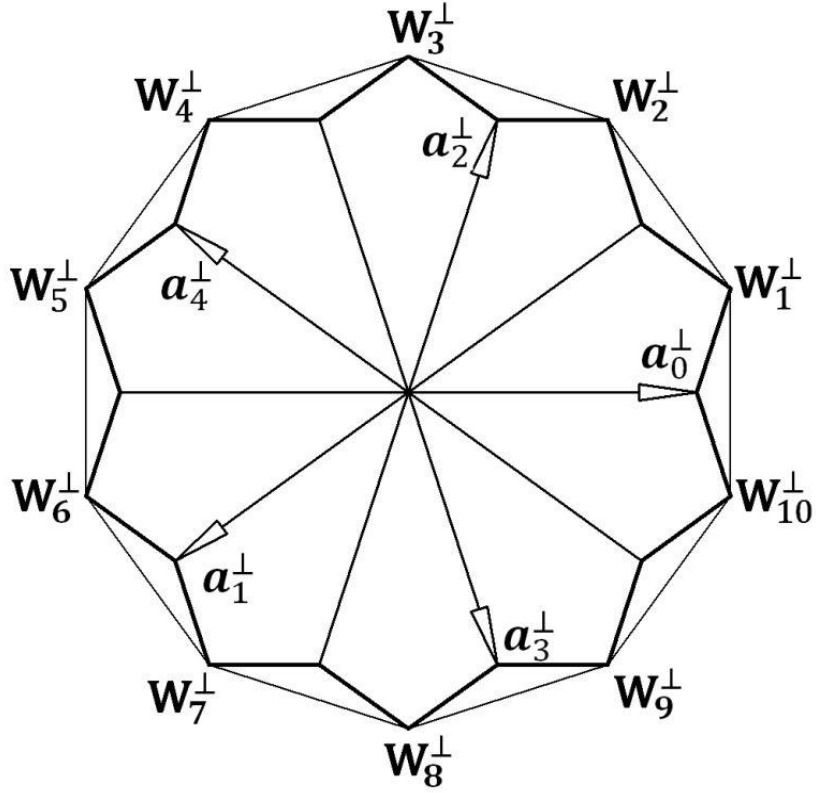

*FIG. 4. Acceptance domain in the shape of a regular ten - arms star. Distance between lines ($\mathbf{W}_1^\perp, \mathbf{W}_3^\perp$) and ($\mathbf{W}_6^\perp, \mathbf{W}_8^\perp$) is equal to $\mathbf{Z}_i^{1\perp}$. Outer vertexes of the star coincide with the vertexes $\mathbf{W}_1$-$\mathbf{W}_{10}$ of the acceptance domain presented in Fig. 2. Interior vertexes coincide with the perpendicular projections of vectors $\pm\mathbf{a}_i^\perp$. The nodes with perpendicular coordinates projected inside a decagon formed by interior vertexes have a smaller value (4) then all their neighbors with respect to star $\mathbf{Z}^1$.*

The next section justifies a fundamental physical basis of our method and demonstrates its relation with Landau crystallization theory.

### 3. Landau theory of crystallization and quasilattices

The pentagonal QL's described before can be also constructed using a generalization of Landau crystallization theory [22-23]. Near a crystallization point the distribution density of structural units in the frame of the theory can be written as

$$\rho(\mathbf{R}) = \rho_0 + \delta\rho(\mathbf{R}), \qquad (5)$$

where $\rho_0$ is an atomic density before crystallization and $\delta\rho(R)$ corresponds to a critical density deviation induced by a quasicrystalline order appearance. According to Landau theory $\delta\rho(R)$ is an irreducible function breaking an isotropic symmetry. The corresponding plane-wave expansion takes the form:

$$\delta\rho(\mathbf{R}) = \sum_k \rho_k \exp(i\mathbf{b}_k \mathbf{R}) \qquad (6)$$

Here **R** is a radius vector, $\rho_k$ is an amplitude of the critical wave with the wave vector $\mathbf{b}_K$. Index *k* runs over ten vectors $\mathbf{b}_K$ rotationally equivalent with respect to $C_{10v}$ symmetry group. The QL free energy *F* can be expanded into a power series of complex amplitudes $\rho_k = |\rho_k|\exp(i\phi_k)$ [24]. The density deviation $\delta\rho(R)$ is real, so that $\rho_k = \rho_m^*$ for $\mathbf{b}_K = -\mathbf{b}_m$. The rotational symmetry leads to equality of all $|\rho_k|$ values: $|\rho_k| = \rho_\Delta/2$. Since *F* is real, it doesn`t depend on 4 mutually orthogonal linear combination of phases $\phi_k$ [24]. Using a phase parameterization [24] in terms of Goldstone variables **u** and **v** (which don`t change the free energy) $\delta\rho(R)$ can be rewritten in the following real form:

$$\delta\rho(\mathbf{R}) = \rho_\Delta \sum_{n=0}^{4} \cos(\mathbf{b}_n \mathbf{R} + \phi_n) = \rho_\Delta \sum_{n=0}^{4} \cos(\mathbf{b}_n \mathbf{R} - \mathbf{b}_n \mathbf{u} - \mathbf{b}_\mathbf{n}^\perp - 2\pi\xi/5) \qquad (7)$$

For the pentagonal QL's under consideration the basis vectors $\mathbf{b}_i$ and value of $\xi$ have the form:

$$\mathbf{b}_n = \frac{4\pi\tau^2}{5a}\left(\cos\left(\frac{n2\pi}{5}\right), \sin\left(\frac{n2\pi}{5}\right)\right) \qquad (8)$$

$$\mathbf{b}_n^\perp = \frac{4\pi}{5a\tau^2}\left(\cos\left(\frac{n6\pi}{5}\right), \sin\left(\frac{n6\pi}{5}\right)\right) \qquad (9)$$

$$\xi = \frac{1}{2\pi}\sum_{n=0}^{4}\phi_n \qquad (10)$$

where *a*=1 is a length of the basis vectors in the (see Eq. (1)) parallel space, $\tau = (\sqrt{5} + 1)/2$ is a golden mean. Note, that Eqs. (8-9) differ from a conventional definition of the reciprocal space basis vectors by self-similarity coefficient $\tau^2$ only. Therefore basis vectors of the critical density waves can be also used as basis vectors of the QL reciprocal space.

Second and fourth-order terms of Landau free energy expansion *F* do not depend on ξ value. This dependence appears starting from a fifth-order term. Therefore, ξ is not Goldstone variable. This term is as follows:

$$F_5 = a^5(\rho_0\rho_1\rho_2\rho_3\rho_4 + \rho_5\rho_6\rho_7\rho_8\rho_9) = a_5/16\rho_\Lambda^5 \cos(2\pi\xi) \qquad (11)$$

where $a_5$ is phenomenological coefficient, $\rho_k = \rho_{k+5}^*$. Contribution (11) to the free energy *F* is the same for all integer ξ values. A more detailed analysis shows that free energy *F* is also the same for all integer ξ values. This explains the structural similarity of QL's different from each other by ξ value only. Note also, that if these QL's correspond to a free energy minimum, the value of $a_5$ coefficient should be negative.

Since the critical density deviation *δρ(R)* gives the main contribution to the density near the crystallisation point, the QL positions should coincide (see Fig. 5) with the highest maxima of Eq. (7). Analogous approach was successfully used in [25].

Panels a) and b) of Fig. 5 demonstrate a correspondence between the maxima of density deviation (7) and positions of QL's shown in figures 1 and 3, respectively. The maxima not filled by the QL nodes are also presented. To distinguish between filled and not filled maxima of *δρ(R)* function the selection rule is necessary. It plays a role of a second local order parameter, not taken into account by conventional theories [22-24].

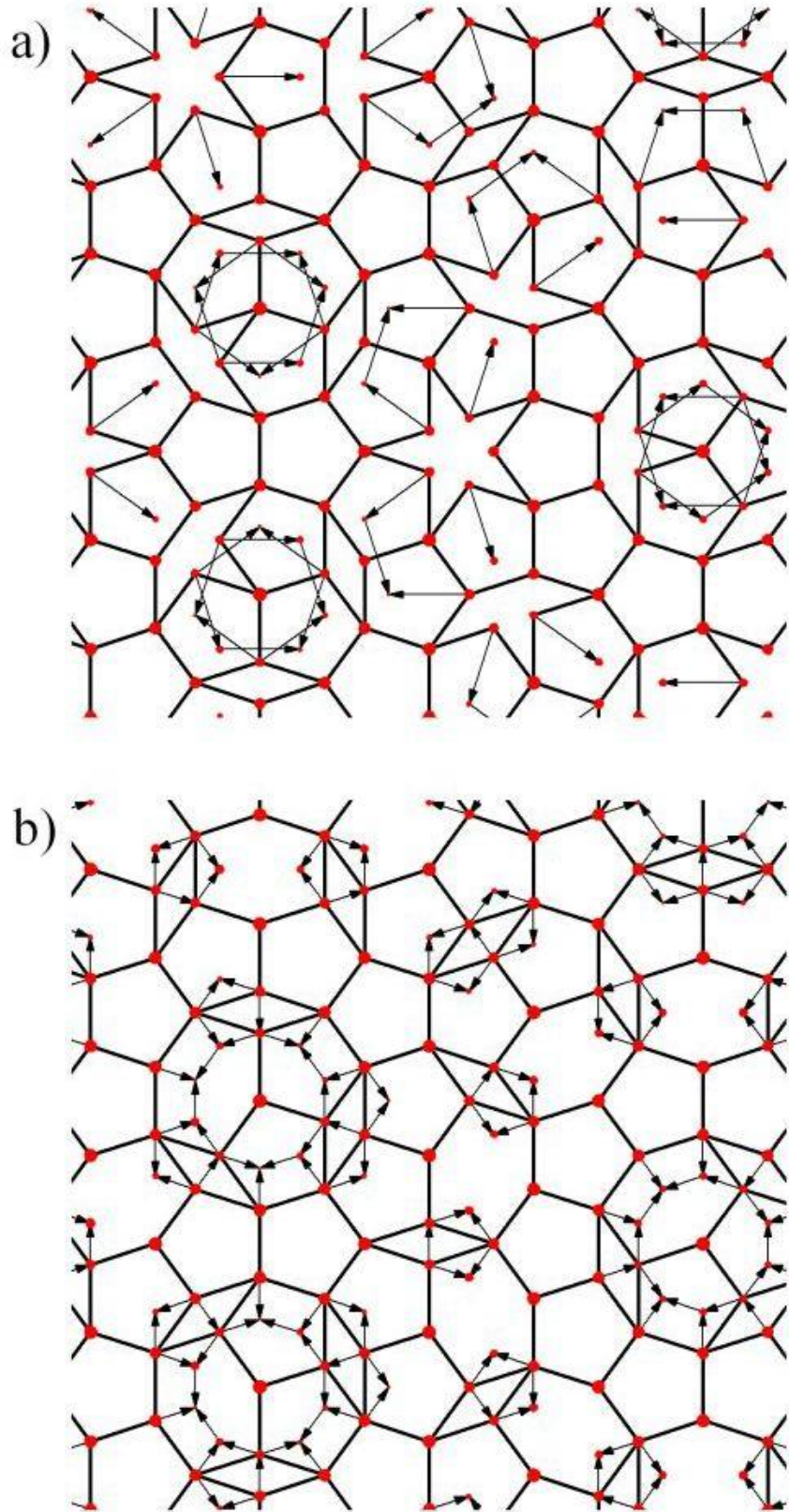

FIG. 5. *A correspondence between the maxima of critical density deviation (7) and the QL positions. Regions with maximal values of the density deviation $\delta\rho(\mathbf{R})>3.5$ are shown in red. All positions of the QL's constructed (pannels a) and b)) coincide with the maxima.*

Adapting the selection rules to the theory of crystallization, we stress two points. 1) A higher maximum of $\delta\rho(\mathbf{R})$ function corresponds to a smaller value of Eq. (4). Therefore, instead of distances (4) we can compare the heights of maxima. To select the filled maxima the local order of neighboring peaks is analyzed. For this purpose we use the irreducible star $\mathbf{Z}$ of translations to the given node from its neighbors (see Section 2). Only the maxima translated into the peak under selection by the parallel projections of the star vectors are considered. For example, in the case 5a) arrows aimed to the excluded peak from the higher ones present $\mathbf{Z}_i^{\parallel}$ projections. The peak exclusion eliminates prohibited tiles (thick rhombuses) and results in ordering of the structure. Finally, the selection rule for the QL shown in Fig. 5(a) can be reformulated the following way: *If two maxima of the function $\delta\rho(\mathbf{R})$ are separated by translation $\mathbf{Z}_i^{\parallel}$, the less intensive (and energetically unfavorable) one should be excluded.* In Fig. 5(b) arrows note parallel components of translations $\mathbf{Z}_i^{1\parallel}$. They are directed to the node under selection from its translational neighbors with the higher value of $\delta\rho(\mathbf{R})$. The node is excluded if at least two arrows are directed to it.

Indeed, conceptions of projection window or atomic surface are important and convenient in crystallography of QC's. However, the equivalent physical theory can be constructed without these purely geometrical objects. As it is clear now, any form of the selection rule serves to eliminate energetically unfavorable local structural configurations. Here, a similarity with the results obtained in the frame of molecular dynamics of QC's exists. An assembly of QC structures by molecular dynamics methods requires two-minima interaction potential between the particles [26]. Two minima are separated by a maximum. Possibly, the corresponding distance correlates with the length of unfavorable translations in our approach.

## 5. Conclusion

We propose a new method to construct quasilattices. In our method the QL positions are obtained projecting integer nodes of N-dimensional space *E*, as well as in the cut and projection methods. However the method proposed does not use conceptions of acceptance domain or atomic surfaces. Nodes included into the QL are selected basing on their local order. This order is characterized by the irreducible translational star **Z** to the node under selection from its neighbors. To select the node *(j)* the values $l_j = |\mathbf{r}_j^\perp + \mathbf{v}|$ and $l_m = |\mathbf{r}_m^\perp + \mathbf{v}|$ are calculated. Here index *(m)* runs over the neighbor nodes, **v** is Goldstone degree of freedom of the tiling, $\mathbf{r}_j^\perp$ is the projection of the node *(i)* upon the perpendicular subspace $E^\perp$. Comparing value $l_j$ with those of $l_m$ the node selection is performed.

The method proposed is based on Landau crystallization theory. Since, in the vicinity of crystallization point the critical density deviation $\delta\rho(\mathbf{R})$ gives the greatest contribution to the density change, the QL positions coincide with the highest maxima of $\delta\rho(\mathbf{R})$. The node fills the maximum of $\delta\rho(\mathbf{R})$, if this maximum is not surrounded by specifically situated higher maxima. All the QL's constructed in this paper in the frames of N-dimensional crystallography can be also deduced basing on Landau crystallization theory.

**Acknowledgments**: The authors would like to thank the graduate student at the Southern Federal University S. Galitskiy for assistance in the execution of the manuscript